\documentclass[a4paper,twoside]{article}
\usepackage{bbm}
\usepackage{amsmath} 
\catcode`\@=11
\long\def\@makefntext#1{
\protect\noindent \hbox to 3.2pt {\hskip-.9pt  
$^{{\eightrm\@thefnmark}}$\hfil}#1\hfill}		

\def\@makefnmark{\hbox to 0pt{$^{\@thefnmark}$\hss}}	
	
\def\ps@myheadings{\let\@mkboth\@gobbletwo
\def\@oddhead{\hbox{}
\rightmark\hfil\eightrm\thepage}   
\def\@oddfoot{}\def\@evenhead{\eightrm\thepage\hfil
\leftmark\hbox{}}\def\@evenfoot{}
\def\sectionmark##1{}\def\subsectionmark##1{}}

\oddsidemargin=\evensidemargin
\newcounter{sectionc}\newcounter{subsectionc}\newcounter{subsubsectionc}
\renewcommand{\section}[1] {\vspace{12pt}\addtocounter{sectionc}{1} 
\setcounter{subsectionc}{0}\setcounter{subsubsectionc}{0}\noindent 
	{\tenbf\thesectionc. #1}\par\vspace{5pt}}
\renewcommand{\subsection}[1] {\vspace{12pt}\addtocounter{subsectionc}{1} 
\setcounter{subsubsectionc}{0}\noindent 
{\bf\thesectionc.\thesubsectionc. {\kern1pt \bfit #1}}\par\vspace{5pt}}
\renewcommand{\subsubsection}[1] {\vspace{12pt}\addtocounter{subsubsectionc}{1}
	\noindent{\tenrm\thesectionc.\thesubsectionc.\thesubsubsectionc.
	{\kern1pt \tenit #1}}\par\vspace{5pt}}
\newcommand{\nonumsection}[1] {\vspace{12pt}\noindent{\tenbf #1}
	\par\vspace{5pt}}

\newcounter{appendixc}
\newcounter{subappendixc}[appendixc]
\newcounter{subsubappendixc}[subappendixc]
\renewcommand{\thesubappendixc}{\Alph{appendixc}.\arabic{subappendixc}}
\renewcommand{\thesubsubappendixc}
	{\Alph{appendixc}.\arabic{subappendixc}.\arabic{subsubappendixc}}

\renewcommand{\appendix}[1] {\vspace{12pt}
        \refstepcounter{appendixc}
        \setcounter{figure}{0}
        \setcounter{table}{0}
        \setcounter{lemma}{0}
        \setcounter{theorem}{0}
        \setcounter{corollary}{0}
        \setcounter{definition}{0}
        \setcounter{equation}{0}
        \renewcommand{\thefigure}{\Alph{appendixc}.\arabic{figure}}
        \renewcommand{\thetable}{\Alph{appendixc}.\arabic{table}}
        \renewcommand{\theappendixc}{\Alph{appendixc}}
        \renewcommand{\thelemma}{\Alph{appendixc}.\arabic{lemma}}
        \renewcommand{\thetheorem}{\Alph{appendixc}.\arabic{theorem}}
        \renewcommand{\thedefinition}{\Alph{appendixc}.\arabic{definition}}
        \renewcommand{\thecorollary}{\Alph{appendixc}.\arabic{corollary}}
        \renewcommand{\theequation}{\Alph{appendixc}.\arabic{equation}}
        \noindent{\tenbf Appendix \theappendixc #1}\par\vspace{5pt}}
\newcommand{\subappendix}[1] {\vspace{12pt}
        \refstepcounter{subappendixc}
        \noindent{\bf Appendix \thesubappendixc. {\kern1pt \bfit #1}}
	\par\vspace{5pt}}
\newcommand{\subsubappendix}[1] {\vspace{12pt}
        \refstepcounter{subsubappendixc}
        \noindent{\rm Appendix \thesubsubappendixc. {\kern1pt \tenit #1}}
	\par\vspace{5pt}}

\newcommand{\textlineskip}{\baselineskip=13pt}
\newcommand{\smalllineskip}{\baselineskip=10pt}

\newcommand{\copyrightheading}[1]
	{\vspace*{-2.5cm}\smalllineskip{\flushleft
	{\footnotesize Quantum Information and Computation, Vol.~1, No.~3 (2001) 79--86 #1}\\
	{\footnotesize \copyright\kern2pt Rinton Press}\\
	 }}

\def\abstracts#1#2#3{{
	\centering{\begin{minipage}{4.5in}\footnotesize\baselineskip=10pt
	\parindent=0pt #1\par 
	\parindent=15pt #2\par
	\parindent=15pt #3
	\end{minipage}}\par}} 

\def\keywords#1{{
	\centering{\begin{minipage}{4.5in}\footnotesize\baselineskip=10pt
	{\footnotesize\it Keywords}\/: #1
	 \end{minipage}}\par}}
\def\communicate#1{{
	\centering{\begin{minipage}{4.5in}\footnotesize\baselineskip=10pt
	{\footnotesize\it Communicated by}\/: #1
	 \end{minipage}}\par}}

\renewenvironment{thebibliography}[1]
        {\frenchspacing
	 \ninerm\baselineskip=11pt
         \begin{list}{\arabic{enumi}.}
        {\usecounter{enumi}\setlength{\parsep}{0pt}     
	 \setlength{\leftmargin 12.7pt}{\rightmargin 0pt}
         \setlength{\itemsep}{0pt} \settowidth
	{\labelwidth}{#1.}\sloppy}}{\end{list}}

\newcounter{itemlistc}
\newcounter{romanlistc}
\newcounter{alphlistc}
\newcounter{arabiclistc}

\def\pmb#1{\setbox0=\hbox{#1}
	\kern-.025em\copy0\kern-\wd0
	\kern.05em\copy0\kern-\wd0
	\kern-.025em\raise.0433em\box0}

\def\fnt#1#2{\footnotetext{\kern-.3em
	{$^{\mbox{\scriptsize #1}}$}{#2}}}

\def\fpage#1{\begingroup
\voffset=.3in
\thispagestyle{empty}\begin{table}[b]\centerline{\footnotesize #1}
	\end{table}\endgroup}

\def\runninghead#1#2{\pagestyle{myheadings}
\markboth{{\protect\footnotesize\it{\quad #1}}\hfill}
{\hfill{\protect\footnotesize\it{#2\quad}}}}
\font\tenrm=cmr10
\font\tenit=cmti10 
\font\tenbf=cmbx10
\font\bfit=cmbxti10 at 10pt
\font\ninerm=cmr9

\font\eightrm=cmr8

\newtheorem{theorem}{\indent Theorem}

\newtheorem{lemma}{Lemma}

\def\qed{\hbox{${\vcenter{\vbox{	          
   \hrule height 0.4pt\hbox{\vrule width 0.4pt height 6pt
   \kern5pt\vrule width 0.4pt}\hrule height 0.4pt}}}$}}

\def\RR{\ensuremath{\mathbbm{R}}} 
\def\CC{\ensuremath{\mathbbm{C}}} 
\def\id{\ensuremath{\mathbbm{1}}} 
\def\LZR{\ensuremath{\mathrm{L}^{\!\rule[-0.5ex]{0mm}{0mm}2}(\RR)}}
\def\cE{{\cal E}}
\def\cH{{\cal H}}
\def\tr{\mathrm{tr}}
\def\ket#1{\left| #1\right>}
\def\bra#1{\left< #1\right|}

\def\qed{\rule{1ex}{1ex}}
\def\assign{\mathrel{\raise.095ex\hbox{:}\mkern-4.2mu=}}

\newcommand{\etal}{\emph{et al.}}
\newcommand{\Eqref}[1]{Eq.~(\ref{#1})}
\newcommand{\Ineqref}[1]{Ineq.~(\ref{#1})}

\newcommand{\Leref}[1]{Lemma~\ref{#1}}
\begin{document}
\setlength{\textheight}{8.0truein}    

\runninghead{Distillability Criterion for Gaussian States} 
            {Giedke, Duan, Zoller, and Cirac }

\normalsize\textlineskip
\thispagestyle{empty}
\setcounter{page}{1}

\copyrightheading{}	

\vspace*{0.88truein}

\fpage{1}
\centerline{\bf
DISTILLABILITY CRITERION FOR ALL BIPARTITE GAUSSIAN STATES}
\vspace*{0.37truein}
\centerline{\footnotesize 
G GIEDKE\footnote{email: geza.giedke@uibk.ac.at}, L-M DUAN,
I CIRAC, AND P ZOLLER,}
\vspace*{0.015truein}
\centerline{\footnotesize\it Institut f\"ur Theoretische Physik, Universit\"at
Innsbruck, Technikerstrasse 25, 6020 Innsbruck, Austria}
\baselineskip=10pt
\vspace*{0.21truein}
\abstracts{
We prove that all inseparable Gaussian states of two modes can be
distilled into maximally entangled pure states by local operations. 
Using this result we show that a bipartite Gaussian state of
arbitrarily many modes can be distilled if and only if its partial
transpose is not positive.}{}{}
\vspace*{10pt}
\vspace*{3pt}
\communicate{S Braunstein and C Fuchs}

\vspace*{1pt}\textlineskip	

\noindent
The existence of \emph{pure} entangled states of two or more systems
entails the possibility of finding new applications of Quantum
Mechanics, in particular in the fields of computation and
communication \cite{entQBapp}.  In practice, however, systems are
exposed to interactions with the environment, that transform pure into
\emph{mixed} states, which may no longer be useful for quantum
communication.  Fortunately, there exist methods to recover pure
entangled states from mixed ones in certain situations. These
processes are called \emph{entanglement distillation} (or
purification) \cite{Ben96}, and consist of local operations and
classical communication transforming several copies of a mixed
entangled state into (approximately) pure entangled states which can
then be used for quantum communication. In fact, applying this method in
the appropriate way one can construct quantum repeaters
\cite{qrepeater} that should allow efficient quantum communication over
arbitrarily long distances even via a noisy channel. 

For this reason it is important to determine whether a
given state is distillable or not. In general, the answer to this
question is not known. At the moment we only have conditions that are
necessary or sufficient for distillability, but not both. Clearly,
only inseparable states can be distilled. Moreover, as shown by
Horodecki \emph{et al.} 
\cite{HoroDist}, there exists a stronger necessary condition, namely
that $\rho$ must have non-positive partial transpose (npt).
In fact, there are entangled states which are not distillable since
their density matrices remain positive under partial transposition
\cite{BE}. Furthermore, there is evidence that this condition is
not sufficient, since there exist npt states that nevertheless seem to
be undistillable \cite{nptBE}. The existence of undistillable npt
states would have interesting consequences such as non-additivity and
non-convexity of the entanglement of formation \cite{nptbe_cons}. 

On the other hand, a useful sufficient criterion,
the so-called reduction criterion \cite{RC}, has been established. It
states that, given a state $\rho$ on the composite Hilbert space
$\cH=\cH_A\otimes\cH_B$, if there exists a vector $|\psi\rangle\in\cH$
such that  
\begin{equation}\label{RC}
\bra{\psi}\tr_B\rho\otimes\id-\rho\ket{\psi}<0.
\end{equation}
then the state $\rho$ is distillable. Here, tr$_B$ stands for the partial
trace with respect to the second subsystem. An important aspect of
this criterion is that if one can find a state $|\psi\rangle$
satisfying (\ref{RC}), then one can explicitly construct a protocol to
distill $\rho$.
 
Up until now, nearly all work on the distillability problem has
considered states of finite 
dimensional systems, see \cite{primer} for a current overview. 
In particular it was shown that states systems consisting of one
qubit and an $N$-level, $N\geq 2$ system are distillable if and only
if (iff) they are npt \cite{QBDist,nptBE}.  
An alternative setting for quantum information processing, which
considers infinite dimensional systems [continuous variables (CV) or
``modes''] in Gaussian states is receiving increasing attention
recently \cite{CVQTel,BraunsteinPati}. For CV systems some distillation
protocols for particular states have been proposed \cite{CVDist}, and
the existence of bound entangled states has been proved
\cite{CVBE,GBE}, but the question of distillability in general has not
been addressed. 

In this article we answer this question completely for all Gaussian
states. We will prove that
\begin{theorem} (Distillability Criterion)\\
\label{Th}
A Gaussian state of $N\times M$ modes is distillable if and only
if its partial transpose is negative.
\end{theorem}

This shows that there are no npt bound entangled Gaussian states and,
in particular, that for systems of $1\times N$ modes all entangled
states (npt is necessary for inseparability of such systems
\cite{GBE,Duan99,Sim99}) are distillable and thus useful for quantum
communication.  Moreover, our proof, which is based in part on the
reduction criterion, provides an explicit protocol that accomplishes
distillation for all those states.
After introducing  the necessary notation and properties of Gaussian
states, the remainder of the paper is devoted to the proof of
Theorem \ref{Th}.  

We consider bipartite systems composed of two subsystems, A and B, 
which consist of $N$ and $M$ ``modes''
[distinguishable infinite dimensional quantum systems with
Hilbert space $\LZR$], respectively. The joint system is
referred to as a ``$N\times M$ system''. It is convenient to describe
the state $\rho$ of such a system by its characteristic function
(e.g., \cite{chfct}) 
\begin{equation}
\chi(x) = {\rm tr} [\rho D(x)]. 
\end{equation}
Here $x = (q_1,p_1,\dots,q_{N+M},p_{N+M})\in\RR^{2N+2M}$ is a real
vector and  
\begin{equation}
D(x) = e^{-i\sum_k (q_k X_k + p_k P_k)},
\end{equation}
where $X_k$ and $P_k$ are operators satisfying the canonical
commutation relations ($\hbar =1$). 
A characteristic function $\chi$ uniquely defines a state
$\rho_\chi$. In the following we exclusively consider \emph{Gaussian
states}, i.e.\ states for which
$\chi$ is a Gaussian function of $x$ \cite{Manu}
\begin{equation}\label{charfct}
\chi(x) = e^{-\frac{1}{4}x^T\gamma x - id^T x}, 
\end{equation}
where $\gamma$ is the \emph{correlation matrix} (CM) 
and $d\in\RR^{2N+2M}$ the \emph{displacement}. Thus,
a Gaussian state is fully characterized by its CM $\gamma$
and displacement $d$.
These states are of particular interest, since they comprise
essentially all CV states that can be prepared in the lab with current
technology.

A matrix $\gamma$ is the CM of a physical state iff
(e.g. \cite{Scu89}) it is strictly positive, real, symmetric
$2(N+M)\times2(N+M)$ and satisfies
\begin{equation}\label{CM}
\gamma\geq J^T\gamma^{-1}J,
\end{equation}
where $J_{N+M} = \bigoplus_{k=1}^{N+M} J_1$ \cite{fn_oplus} with
$J_1 = \left(\begin{array}{cc} 0&-1\\1&0 \end{array} \right)$.

Simon \cite{Sim99} noted that for CV states
partial transposition is equivalent to the orthogonal transformation 
$\Lambda_B(q_A,p_A,q_B,p_B)=(q_A,p_A,q_B,-p_B)$ on phase space, i.e.,
the momentum coordinates referring to B are inverted. 
For a Gaussian state this means that its CM is changed to
$\tilde\gamma = \Lambda_B\gamma\Lambda_B$ and the displacement to
$\Lambda_Bd$.  A Gaussian state with CM $\gamma$ has
negative partial transpose (npt) iff $\tilde\gamma$ does not satisfy
\Ineqref{CM} \cite{Sim99,GBE}, or, equivalently, iff
\begin{equation}\label{Gnpt}
\gamma \not\geq \tilde J^T\gamma^{-1}\tilde J,
\end{equation}
where $\tilde J = \Lambda_B J \Lambda_B^T$ is the ``partially
transposed'' $J$ in which the $J_1$'s corresponding to B's modes are
replaced by $-J_1$.

The first part of the proof of the theorem is concerned with the
special case of a bipartite two-mode Gaussian state: 
$N=M=1$. Any such state can be transformed into what we called the
{\em standard form}, using local unitary operations only
\cite{Duan99,Sim99}. For a state in standard form the displacement
$d=0$ and the CM $\gamma$ has the simple form 
\begin{equation}\label{corrmat} 
\gamma  =  \left( \begin{array}{cc} A&C\\C^T&B 
\end{array} \right), 
\end{equation} 
where\\
\begin{equation}\label{stdform} 
A = \left( \begin{array}{cc} n_a&0\\0&n_a 
\end{array} \right)\!,\,\, B = \left( \begin{array}{cc} n_b&0\\0&n_b 
\end{array} \right)\!,\,\, C = \left( \begin{array}{cc} k_x&0\\0&k_p 
\end{array} \right). 
\end{equation}
The local unitaries needed to achieve this form are linear
Bogoliubov transformations, i.e., generated by Hamiltonians that are
at most quadratic in the operators $X_{1,2},P_{1,2}$
The four real parameters $(n_a,n_b,k_x,k_p)$ fully characterize a 
$1\times1$ Gaussian state up to local linear Bogoliubov transformations (LLBT). 
They can be easily calculated from the four LLBT-invariant 
determinants $\det A, \det B, \det C$, and 
$\det\gamma$ as follows: 
\begin{subequations}\label{param} 
\begin{equation} 
n_a = \sqrt{\det A}, 
n_b = \sqrt{\det B}, 
k_xk_p = \det C, 
\end{equation} 
\begin{equation}
(n_an_b-k_x^2)(n_an_b-k_p^2) = \det\gamma.
\end{equation} 
\end{subequations} 
Without loss of generality we choose $k_x\ge|k_p|$.  
We call a state \emph{symmetric}, if $n_a=n_b=n$, or, equivalently, if
$\det A=\det B$. 

Now we are prepared for the proof of Theorem \ref{Th}.
We state the three main steps of the proof in three lemmas, which we
prove in the remainder of this article. 

\begin{lemma} (Distillability of Symmetric $1\times1$ States)\\
\label{symmDist}
A symmetric $1\times1$ Gaussian state with non-positive partial
transpose is distillable. 
\end{lemma}

\begin{lemma} (Symmetrization of $1\times 1$ States)\\
\label{symm}
Every $1\times1$ Gaussian state with non-positive partial
transpose can be locally transformed into a symmetric npt state. 
\end{lemma}

\begin{lemma} (Concentrating Inseparability in two Modes)\\
\label{conc}
Every $N\times M$ Gaussian state with non-positive partial
transpose can be locally transformed into a $1\times1$ npt state.  
\end{lemma}

\noindent\textbf{Proof of Theorem \ref{Th}: } The ``only if''-part of the
Theorem was proven proven by the Horodeckis in  \cite{HoroDist}. 
The ``if''-part is clearly implied by these three
Lemmas, since by \Leref{conc} the $N\times M$ case can be reduced to
the $1\times1$ case, and that case by \Leref{symm} to the symmetric
case. 
\hfill\qed

For the proof of Lemmas \ref{symmDist} and \ref{symm}, it is useful to
re-express the conditions (\ref{CM},\ref{Gnpt}) for $1\times1$ states
in terms of the parameters (\ref{param}). We find that $\gamma$ is CM
of a physical state iff
\begin{subequations}\label{stdphys} 
\begin{eqnarray} 
\label{stdphysa} 
(n_an_b-k_x^2)(n_an_b-k_p^2)+1&\geq& n_a^2+n_b^2+2k_xk_p,\\ 
\label{stdphysb} 
n_an_b-k_x^2 &\geq& 1, 
\end{eqnarray} 
\end{subequations}    
and that $\gamma$ is CM of an inseparable (or, equivalently, npt)
state, iff in addition it holds that  
\begin{eqnarray}\label{insep} 
(n_an_b-k_x^2)(n_an_b-k_p^2)+1 &<&n_a^2+n_b^2-2k_x k_p. 
\end{eqnarray} 

\noindent\textbf{Proof of \Leref{symmDist}: }
For this we use that a state is distillable,
if there exists a pure state $\left|\psi\right\rangle$ such that
\Ineqref{RC} holds. This condition was proved in \cite{RC} to be
sufficient for distillability of finite dimensional systems. 
Its extension to infinite dimensions is straightforward: and proved in
the appendix.

We show now that for any symmetric npt Gaussian state $\rho$
\Ineqref{RC} is satisfied with $\ket{\psi}$ taken as the pure two-mode
squeezed state $\ket{\psi}=\frac{1}{\cosh r}\sum_n\tanh^nr\ket{nn}$
for sufficiently large $r>0$. Note that $\ket{\psi}$ is a symmetric
Gaussian state in standard form. We denote its CM by $\gamma_\psi$ and
the four parameters (\ref{param}) are $n_a=n_b=\cosh 2r,
k_x=-k_p=\sinh 2r$. Let $\gamma_\rho$ denote the CM of $\rho$.
With these choices, \Ineqref{RC} becomes \cite{Scu98}
\begin{equation} 
2\left[ \det(\gamma_{\tr_B\rho}+\gamma_{\tr_B\psi}) 
\right]^{-1/2}-4\left[ \det(\gamma_\rho+\gamma_\psi) 
\right]^{-1/2}<0.  
\end{equation}    
In the limit of large $r$ (keeping only the leading terms in $e^r$)
this becomes after some simple algebra
\begin{equation}\label{RCsymm} 
(n-k_x)(n+k_p)<1.
\end{equation} 
But Ineq.\ (\ref{RCsymm}) is implied by the
inseparability criterion for symmetric states: if $n_a=n_b=n$ then
\Ineqref{insep} simplifies to 
\begin{equation}\label{symmcond}
|n^2-k_xk_p-1|<n(k_x-k_p). 
\end{equation}
For inseparable states we observe \cite{Sim99} that $ k_xk_p<0$, which
together with \Ineqref{stdphysb} implies that the LHS of
Ineq. (\ref{symmcond}) is equal to $n^2-k_xk_p-1$ which can be
transformed to $(n-k_x)(n+k_p)+n(k_x-k_p)-1$
from which \Ineqref{RCsymm} follows immediately.\hfill\qed 

Since the local operation that will be shown to achieve symmetrization
involves a measurement, it is more convenient to describe the state here by
its \emph{Wigner function} \cite{chfct}. It is related to the
characteristic function by symplectic Fourier transformation and thus
is Gaussian for Gaussian states. The Wigner CM $\gamma_W$ is related
to the (characteristic) CM by $\gamma_W=J^T\gamma^{-1} J$. We denote
the four LLBT-invariant parameters for the Wigner CM [which are
defined as in (\ref{param})] by $(N_a,N_b,K_x,K_p)$. We use the
following easily checked facts: just as the standard form of $\gamma$,
the standard form of $\gamma_W$ can be obtained by LLBTs. A state is
symmetric iff $N_a=N_b$. The conditions (\ref{stdphys},\ref{insep})
can be formulated equivalently in terms of the parameters
$(N_a,N_b,K_x,K_p)$. While (\ref{stdphysa},
\ref{insep}) are identical for the Wigner parameters, in
(\ref{stdphysb}) ``$\geq$'' is changed to ``$\leq$''. We refer to
these conditions for the Wigner parameters as (\ref{stdphys}W,
\ref{insep}W) in the following.

\noindent\textbf{Proof\ of \Leref{symm}: }
If the state is not symmetric, it means that the reduced state at one
of the two sides has larger entropy than the other. This suggests to
let a pure state interact with the ``hotter'' side to cool it
down. This must be done without destroying the entanglement of
$\rho$. We proceed as follows: $\rho$ is transformed to its
Wigner standard form with parameters $(N_a,N_b,K_x,K_p)$.  
Now assume that $N_b<N_a$, i.e., B is the hotter side
\cite{fn_hot}. Take an ancilla 
mode in the vacuum state and couple it to B's mode by a beam splitter
\cite{fn_ops} 
with transmittivity $\cos^2\theta$.  After a measuring
the ancilla's $X$-operator \cite{fn_meas} results a state $\tilde\rho$
with Wigner CM $\tilde \gamma_W$ of the form (\ref{corrmat}) with
\[\tilde A = \frac{1}{\nu}\left( \begin{array}{cc} 
c^2N_a+s^2D_x & 0\\ 0&c^2N_a+s^2N_aN_b
\end{array} \right),\]  
\[\tilde B = \frac{1}{\nu}\left( 
\begin{array}{cc} N_b&0\\ 0&[c^2N_b+s^2]\nu \end{array}\right),\,\,  
\tilde 
C = \frac{1}{\nu}\left( \begin{array}{cc} cK_x&0\\0&cK_p\nu
\end{array} \right), \] 
where the abbreviations $c=\cos\theta, 
s=\sin\theta, \nu = s^2N_b+c^2$, and $D_{x,p}=N_aN_b-K_{x,p}^2$ were 
used. The condition for symmetry,  
$\det\tilde A| = \det\tilde B|$, requires
\begin{equation}\label{theta} 
\tan^2\theta=\frac{N_a^2-N_b^2}{N_b-D_xN_a}. 
\end{equation} 
Checking (\ref{insep}W) for $\tilde \gamma_W$ one sees that the
inequality is just multiplied by $(N_b\tan^2\theta+1)^{-1}>0$;
therefore the transformed state is inseparable iff the original one
was inseparable. It remains to show that there always exists a
$\theta$ to satisfy 
(\ref{theta}), i.e., that the right hand side of Eq. (\ref{theta}) is
positive. The numerator is positive since we have chosen $N_b<N_a$,
the denominator is positive since $N_b<N_a$ and the second part of
condition (\ref{stdphys}W) imply that $(N_a-D_xN_b)>0$ and the first
part of (\ref{stdphys}W) assures that
$(N_a-D_xN_b)(N_b-D_pN_a)\geq(N_aK_x+N_bK_p)^2\geq0$, hence all
Gaussian states in Wigner standard form can be symmetrized this way.
But since every Gaussian 
state can be brought into Wigner standard form by local unitaries,
this completes the proof of \Leref{symm}.\hfill\qed

To finish the proofs, we now turn to the general case of $N\times M$
modes. Let $\gamma$ be the CM of a npt state. 

\noindent\textbf{Proof of \Leref{conc}: } 
The condition (\ref{Gnpt}) is equivalent to $\gamma\not\geq i\tilde J$
\cite{GBE}. Hence, for every npt state with CM $\gamma$ there exists
a vector $z\in\CC^{2(N+M)}$ such that for some $\epsilon>0$
\begin{equation}\label{npt_cond} 
z^\dagger (\gamma-i\tilde J)z\leq -\epsilon<0. 
\end{equation} 
The idea of the proof is that $\gamma$ can be locally transformed such
that at both sides all but one mode can be discarded, and the 
resulting (reduced) $1\times1$ state is still npt. Then it is
distillable by Lemmas \ref{symmDist} and \ref{symm}.

Write $z$ in \Eqref{npt_cond} as $z=z^{(A)}\oplus z^{(B)}$ with real
and imaginary parts $z^{(x)}_r, z^{(x)}_i$, ($x=A,B$).
We can always find a $z$ such that $z^{(x)}_r$ and $z^{(x)}_i$ are
not skew-orthogonal, i.e.\ $(z^{(x)}_r)^TJz^{(x)}_i\not=0$ for both
$x=A, B$ \cite{fn_notlinind}. 

Now we have to find two symplectic transformations $S_x, x=A,B$
that map the span of $\{ z^{(x)}_r,z^{(x)}_i \}$ to the span of
$\left\{ e_1, e_2\right\}$, where $e_1=(1,0,\dots,0)$ etc. After
performing the local transformation $S=S_A\oplus S_B$, both A and B
can discard all 
but the first mode of their systems and still have an npt entangled (and
thus distillable) state.  That such symplectic transformations $S_A,
S_B$ always exist is seen as follows: 
let $f_1=z_r, f_2= -z_i/(z_r^TJz_i)$; we can always extend ${f_1,f_2}$
into a \emph{symplectic basis} \cite{Arnold} ${f_k:k=1,\dots,2n}$ such
that $f_{2k}^TJf_{2l+1}=\delta_{kl},
f_{2k}^TJf_{2l}=0=f_{2k+1}^TJf_{2l+1}$. Then $S$ defined by $Se_k=f_k$
is symplectic \cite{Arnold} and $S^{-1}$ maps span$\{z_r,z_i\}$ to
span$\{e_1,e_2\}$, i.e.\ 
\begin{equation}\label{firstmode} 
\hat z^{(x)}\assign
S_x^{-1}z^{(x)}=a_xe_1+b_xe_2,
\end{equation} 
$a_x,b_x\in\CC$. Consequently we have for
$\hat\gamma\assign (S_A\oplus S_B)^T\gamma 
(S_A\oplus S_B)$ 
\begin{equation}\label{gtilde} 
(\hat z^{(A)}\oplus\hat z^{(B)})^\dagger(\hat\gamma-i\tilde
J)(\hat z^{(A)}\oplus\hat z^{(B)})<0.
\end{equation} 
Using \Eqref{firstmode} we see that only the matrix elements
$(\hat\gamma)_{kl}$ with $k,l = 1,2, N+1, N+2$ contribute to the lhs
of \Eqref{gtilde}. Thus \Ineqref{gtilde} does not change if we replace
$\hat\gamma$ by the two-mode CM $\hat\gamma_{red}$ obtained from
$\hat\gamma$ by discarding all rows and columns referring to modes
other than 1 and $N+1$. This is the CM of the state in which A and B
discard all but their first mode each. 
Thus \Ineqref{gtilde} shows that the state
$\rho_{red}$ corresponding to $\hat\gamma_{red}$ is npt. But
$\rho_{red}$ is a two-mode state and thus distillable by the first
part of the proof. \hfill\qed

Note that all the operations needed to transform a general $N\times M$
npt state into a symmetric $1\times1$ entangled state can be
implemented quantum optically with current technology: they require
nothing but squeezers, beam splitters, phase shifters \cite{fn_ops}, homodyne
measurements, and the discarding of subsystems. Once a state has been
transformed to symmetric standard form, the protocol of Ref. \cite{RC}
can be used to obtain maximally entangled states in a finite
dimensional Hilbert space.  While a \emph{practical} distillation
protocol for such Gaussian states remains to be found (see however
\cite{CVDist}), it is worth noting, that the main part of the
universal protocol of \cite{RC}, namely the filtering operation and
the joint measurement, are for symmetric Gaussian states implemented
by the procedure of Duan \etal \cite{CVDist}; for details see
\cite[ch. II.8]{BraunsteinPati}.   

In conclusion, we have answered the distillability question for all
Gaussian states: such states are distillable if and only if they are
npt. In particular, all entangled Gaussian states of $1\times N$ modes
are distillable, and there exist no npt bound entangled Gaussian
states.

\nonumsection{Acknowledgements}
\noindent G.G. acknowledges financial support by the
Friedrich-Naumann-Stiftung. This work was supported by the Austrian
Science Foundation (SFB ``Control and Measurement of Coherent Quantum
Systems'', Project 11), the EU (TMR network ERB--FMRX--CT96--0087 and
the project EQUIP, contract IST-1999-11053), the ESF, and the
Institute for Quantum Information GmbH, Innsbruck. 


\nonumsection{References}

\appendix
\noindent We show that Ineq. (\ref{RC}) implies distillability even for
dim$\cH=\infty$. Let $\{\ket{k}:k=0,1,...\}$ be an orthonormal basis
of $\cH$, let 
$\cH_n=\mathrm{span}\{\ket{0},\ket{1},...,\ket{n}\}$, $P_{\cH_n}$
the orthogonal projector on $\cH_n$, and let $\rho$
be a density matrix on $\cH\otimes\cH$. Let
$\cE(\rho)=\tr_B\rho\otimes\id-\rho$ be the map occuring on the
lefthand side of (\ref{RC}). Assume that
$\exists\ket{\psi}\in\cH,\epsilon>0$ such that
$\bra{\psi}\cE(\rho)\ket{\psi}<-\epsilon<0$. Since
$\rho_n=P_{\cH_n}\rho P_{\cH_n}$ converges in the weak topology to
$\rho$, there is $N\geq0$ such that
$\bra{\psi}\cE(\rho_n)\ket{\psi}<-\epsilon/2$ for 
all $n\geq N$. Thus $\rho$ can be projected by local operations onto a
distillable state $\rho_N$ and is therefore itself
distillable.\hfill\qed


\begin{thebibliography}{99}
\bibitem{fn_oplus}
For convenience we use direct sum notation for matrices and vectors.
That is, if $A$ and $B$ are $n\times n$ and $m\times m$ matrices,
resp., then $A\oplus B\in M_{n+m, 
n+m}$ is a block diagonal matrix of blocks $A$ and $B$. Similarly,
if $f_1\in \RR^n$ and $f_2\in \RR^m$ are two vectors, then $f_1\oplus
f_2 \in \RR^{n+m}$ is a vector whose first $n$ components are given by the
entries of $f_1$ and the last $m$ by those of $f_2$.
\bibitem{fn_ops} We use quantum optical terminology for the following
LBT: a beam splitter of transmittivity $t$ coupling the modes $k$ and
$l$ refers to the unitary generated by $t(P_kX_l-X_lP_k)$, a phase
shifter for the $k$th mode is generated by $X_k^2+P_k^2$, a squeezing
transformation by $X_kP_k+P_kX_k$, and a displacement by $aX_k+bP_k$.
\bibitem{fn_meas} If a measurement of an observable $X_k$ or
$P_k$ is performed, the Wigner function of the final state of the
remaining modes is calculated as follows: replace the phase space
coordinate referring to the measured operator by the measured result
and integrate out the conjugate phase space coordinate. Note that the
CM of final state is independent of the measurement outcome.
\bibitem{fn_notlinind}
 In case the original vectors $z_r, z_i$ are skew-orthogonal,
we can replace $z_i$ by $z_i'=z_i+\delta Jz_r$ where $\delta$ is so
small that \Ineqref{npt_cond} still holds for $z'=z_r+iz_i'$.
\end{thebibliography}
\end{document}